\documentclass[aps
]{revtex4}
\usepackage{graphicx}   
\usepackage{subfigure}  

\def\bea{\begin{eqnarray}}
\def\eea{\end{eqnarray}}

\def\sqr#1#2{{\vcenter{\vbox{\hrule height.#2pt
      \hbox{\vrule width.#2pt height#1pt \kern#1pt
         \vrule width.#2pt}
      \hrule height.#2pt}}}}

\begin{document}
\title{Measurement of stimulated Hawking emission in an analogue system
  }

\author{Silke Weinfurtner*, Edmund W. Tedford**, Matthew C. J. Penrice*,
William G. Unruh*, and Gregory A. Lawrence**
}
\affiliation{
*Department of Physics and Astronomy, University of British Columbia, Vancouver, Canada V6T 1Z1\\~\\
**Department of Civil Engineering, University of British Columbia, 6250 Applied
Science Lane, Vancouver, Canada V6T 1Z4.
}

~

~

\begin{abstract}

There is a mathematical analogy between the propagation of fields in a general
relativistic space-time and long (shallow water) surface waves on moving
water.  Hawking argued that black holes emit thermal radiation via a quantum
spontaneous emission.  Similar arguments predict the same effect near wave
horizons in fluid flow. By placing a streamlined obstacle into an open channel
flow we create a region of high velocity over the obstacle that can include
wave horizons.  Long waves propagating upstream towards this region are
blocked and converted into short (deep water) waves.  This is the analogue of
the stimulated emission by a white hole (the time inverse of a black hole),
and our measurements of the amplitudes of the converted waves demonstrate the
thermal nature of the conversion process for this system.  Given the close
relationship between stimulated and spontaneous emission, our findings attest
to the generality of the Hawking process.
\end{abstract}

\maketitle
\section{Introduction}

One of the most striking findings of general relativity is the prediction of
black holes, accessible regions of no escape surrounded by an event horizon.
In the early 70s, Hawking suggested that black holes evaporate via a quantum
instability [1-3]. The study of classical and quantum fields around black
holes shows that a pair of field excitations at temporal frequency $f$ are
created, with positive and negative norm amplitudes $\alpha_f$, $\beta_f$  (Bogoliubov coefficients) related by, 
\bea
{\vert\beta_f\vert^2\over \vert\alpha_f|^2}=\exp\left({-4\pi^2f\over
g_H}\right)
\eea
where $g_H$ is the surface gravity of the black hole [1-4]. Positive norm modes are emitted, while
negative ones are absorbed by the black hole, effectively reducing its mass.
The surface gravity for a non-rotating black hole with a mass $M$ is given by
$g_H
= 1.0×1035/M [{\rm kg/s}]$. Equation (1) is applicable for both stimulated and
spontaneous emission, and at regimes where the quantum physics is dominant. A
comparison of (1) with the Boltzmann-distribution allows one to associate a
temperature $T$ with the black hole, 
\bea
T=1.2\cdot 10^{-12} g_H[{\rm K}]=6.0\cdot 10^{-8} {M_\circ\over M}[{\rm K}]
\eea
Here $M_\circ$ is a solar mass, and the smallest observed black holes are of
this order. Thus black hole evaporation is clearly difficult to observe
directly [5].

In 1981 Unruh showed [6] (see also [7,8]), that there is a mathematical
analogy between the behaviour of classical and quantum fields in the vicinity
of black hole horizons and sound waves in trans-sonic fluid flows. In 2002 it
was argued that surface waves on an open channel flow with varying depth are
an ideal toy model for black hole experiments [9]. The 1981 paper raised the
possibility of doing experiments with these analogues. A difficulty with
Hawking's derivation is its apparent reliance on arbitrarily high frequencies
(the trans-Planckian problem [10-13]). The dispersion relation of gravity
waves creates a natural physical short wavelength cutoff, which obviates this
difficulty. Thus the dependence of the Hawking effect on the high-frequency
behaviour of the theory can be tested in such analogue experiments [14]. While
numerical studies indicate that the effect is independent of short-wavelength
physics, experimental verification of this would strengthen our faith in the
process. The presence of this effect in our physical system, which exhibits
turbulence, viscosity, flow separation, and non-linearities, would indicate
the generic nature of the Hawking thermal process.

\section{\bf Fluid surface gravity waves}

The excitation spectrum of gravity waves on a slowly varying background flow
is well understood and has a dispersion relation given by, 
\bea
f^2=\left({gk\over 2\pi}\right) \tanh(2\pi k h)
\eea
with the frequency, $f = 1/\tau$, where $\tau$ is the wave period; the
wavenumber, $ k =
1/\lambda$, where $\lambda$ is the wavelength; $g$ is the gravitational
acceleration, and $h$ the
depth of the fluid. We neglect surface tension and viscosity. We classify
waves according to the value of $2\pi kh$. For $2\pi  kh < 1$ the dispersion relation
can be approximated by $f = \sqrt{(gh)}k$. These shallow water waves have both
group and phase speed approximately equal to $\sqrt{gh}$.  For $2\pi kh > 1$, the
dispersion relation is approximated by  $f = \sqrt{gk/2\pi}$. The group speed of
these deep water waves is approximately half the phase speed, and both vary as
the square root of the wavelength. For a given water depth, both the group and
phase speeds of deep water waves are less than the group and phase speeds of
shallow water waves.

\begin{figure}
\begin{center}
\includegraphics[width=80mm]{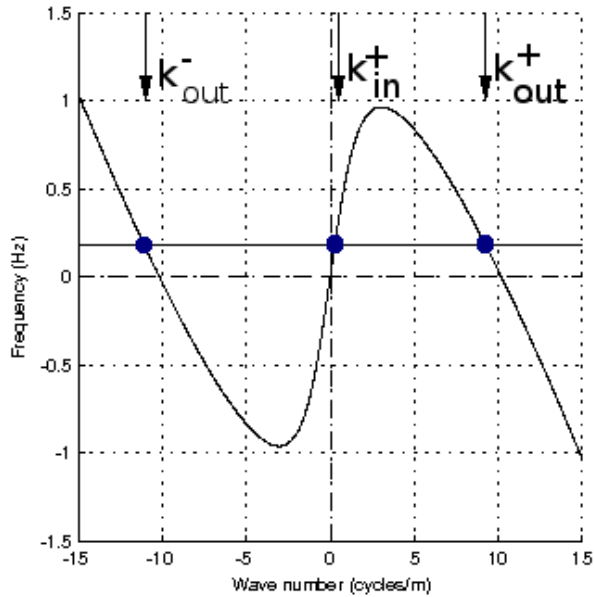}
\caption{{\bf  | Conversion process}.  Dispersion relation for waves
propagating against a flow typical of our experiments. A shallow water wave,
$k_{in}$, sent upstream, is blocked by the flow and converted to a pair of deep
water waves ($k^+_{out}$  and $k^-_{out}$) that are swept downstream. 
}

\end{center}
\end{figure}

In [9] Schützhold and Unruh argued that the equation of motion of shallow
water waves can be cast into a wave equation on a curved spacetime background
if the speed of the background flow varies. Assuming a steady, incompressible
flow the velocity
\bea
v(x)={q\over h(x)}
\eea
Here the two-dimensional flow rate q is fixed. The dispersion relation in the
presence of a non-zero background velocity becomes,
\bea
(f+vk)^2=\left({gk\over 2\pi}\right) \tanh(2\pi kh)
\eea
In Fig.\ 1, the dispersion relation is plotted for a flow typical of our
experiments. Only the branch corresponding to waves propagating against the
flow is plotted. For low frequencies, there are three possible waves, which we
denote according to wavenumber. The first, $k^+_{in}$,  is a shallow water wave with
both positive phase and group velocities, and corresponds to the wave that we
generate in our experiments. The second, $k^+_{out}$, has positive phase velocity,
but negative group velocity. Both waves, $k^+_{in}$ and $k^+_{out}$, are on the positive
norm branch of the dispersion relation. The third, $k^-_{out}$, has both negative
phase and group velocities, and it lies on the negative norm branch. In our
experiment, generated shallow water waves move into a region where they are
blocked by a counter-current, and converted into the other two waves. The goal
of our experiment was the measurement of the relative amplitudes of the
outgoing positive and negative norm modes to test the validity of (1).
(Further conversion from deep-water waves to capillary waves [15,16] are also
possible but are not studied here.)

	The conversion from shallow water to deep water waves occurs where a
counter-current become sufficiently strong to block the upstream propagation
of shallow water waves [16-19]. It is this that creates the analogy with the
white hole horizon in general relativity. That is, there is a region 
that the shallow water waves cannot access, just as light cannot enter a white
hole horizon. Note that while our experiment is on white hole horizon
analogues, it is because they are equivalent to the time inverse of black hole
analogues that we can apply our results to the black hole situation.

\section{Experimental Procedure}

\begin{figure}
\begin{center}
\includegraphics[width=80mm]{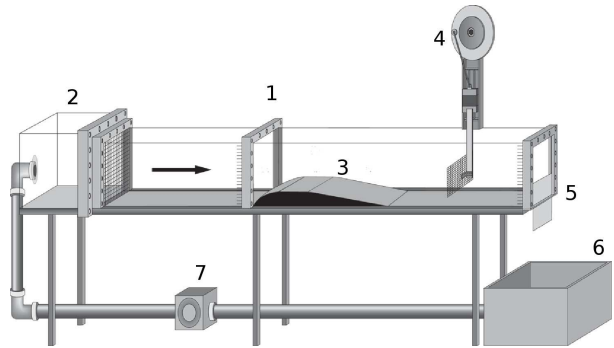}
\caption{{\bf  | Experimental apparatus.} The experimental apparatus used in
our experiments: (1) flume, (2) intake reservoir, (3) obstacle, (4) wave
generator, (5) adjustable weir, (6) holding reservoir, and (7) pump and pump
valve.
}
\end{center}
\end{figure}

Our experiments were performed in a 6.2 m long, 0.15 m wide and 0.48 m deep
flume (Fig.\ 2), and were partly motivated by experiments in similar flumes
[15-20]. We created a spatially varying background flow by placing a 1.55 m
long and 0.106 m high obstacle in the flume. This obstacle was modelled after
an airplane wing with a flat top and a maximum downstream slope of 5.2 degrees
designed to prevent flow separation. We used particle image velocimetry [21]
to determine q, and to verify the absence of flow separation. Shallow water
waves of approximately 2 mm amplitude were
generated 2 m downstream of the obstacle, by a vertically oscillating mesh,
which partially blocked the flow as it moved in and out of the water. The
intake reservoir had flow straighteners and conditioners to dissipate surface
waves produced by the ingoing flow. The flume was transparent to allow
photography through the walls, and the experimental area was covered to
exclude exterior light.

\begin{figure}
\begin{center}
\subfigure{
\includegraphics[width=80mm]{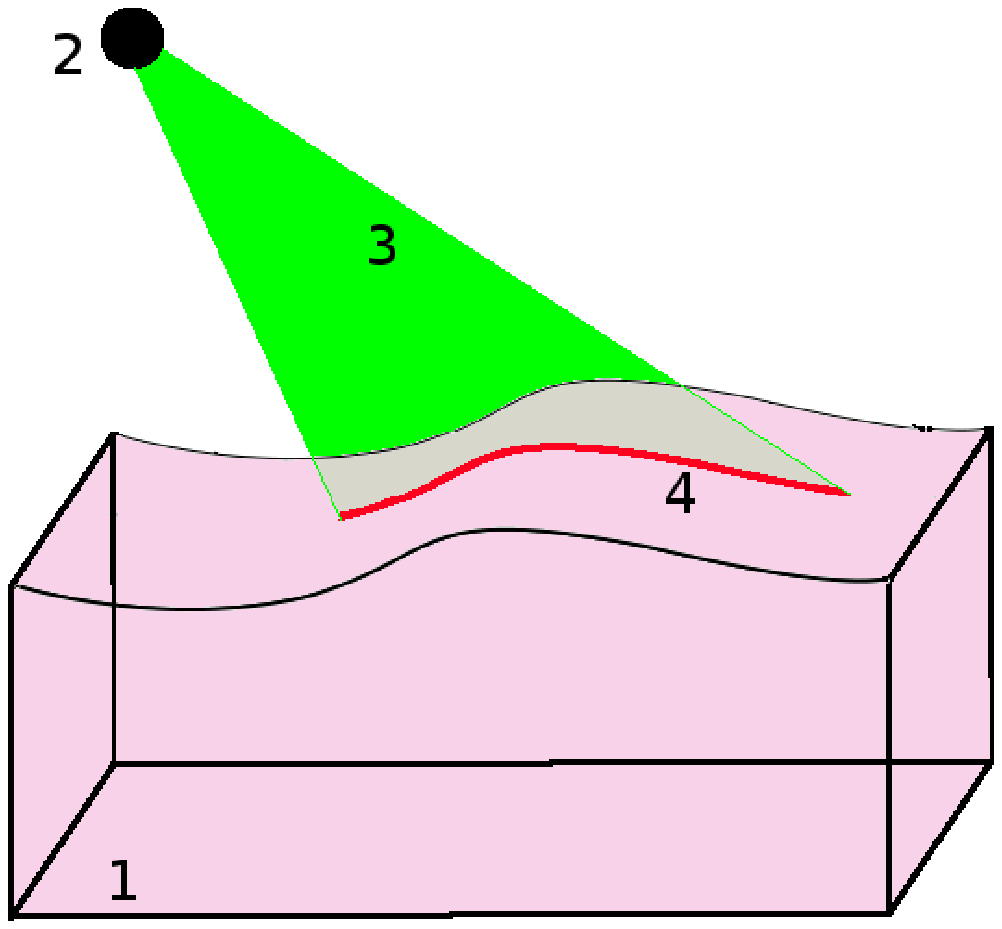}
}
\subfigure{
\includegraphics[width=80mm]{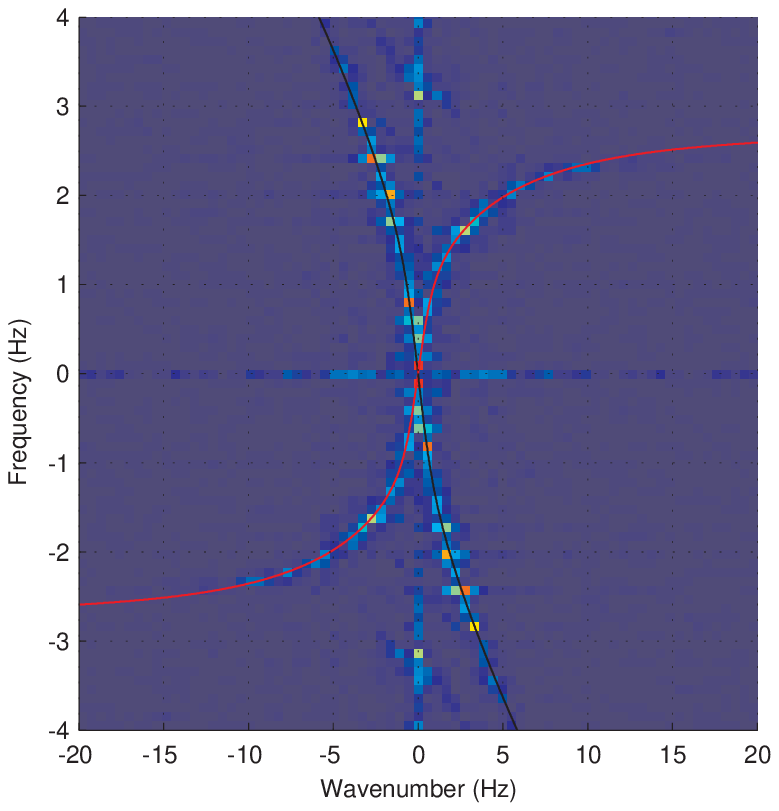}
}
\caption{{\bf  | Surface wave detection. a}, Diagram of light-sheet
projection for surface wave detection: (1) water with dye, (2) Powell lens,
(3) light sheet, and (4) fluorescing water surface.{\bf b}, Fourier transform of
water surface in flat bottom flume without waves; $q = 0.039$ m${}^2$/s and $h =
0.24$ m. Fluctuations lie on upstream (red line) and downstream (black line)
branches of the dispersion relation. Just visible at $f = \pm 3.1$ Hz, $k = 0$ are the
second transverse mode branches of the dispersion relation. Off the dispersion
curves, the background noise amplitudes are less than 0.1mm.
}
\end{center}
\end{figure}

We measured and analysed the variations in water surface height using
essentially the same techniques as in [22]. The water surface was illuminated
using laser-induced fluorescence, and photographed with a high-resolution
(1080p) monochrome camera.  The camera was set up such that the pixel size was
1.3 mm, the imaged area was 2 m wide and 0.3 m high, and the sampling rate was
20 Hz. The green (532 nm) 0.5 W laser light passed through a Powell lens to 
create a thin ($\approx 2$ mm) light sheet (Fig.\ 3a).
Rhodamine-WT dye was dissolved in the water, which fluoresced to create a
sharp ($< 0.2$ mm) surface maximum in the light intensity. We interpolated the
intensity of light between neighbouring pixels to determine the height of the
water surface to subpixel accuracy.

To determine the ambient wave noise in our facility, and to check the
effectiveness of our procedures, we conducted an experiment without the
obstacle in place and with no wave generation. The space and time Fourier
transform of the noise match the dispersion relation for this flow ($q = 0.039
m^2/s$ and $h = 0.24 m$) extremely well (Fig.\ 3b). In general, the amplitude of
the Fourier components has a noise level of less than 0.2 mm away from the
dispersion curves. The apparently elevated noise energy crossing the k axis at
$f = ± 3.1$ Hz is due to the second transverse mode branch of the dispersion
relation (the first transverse mode has a node at the location of the light
sheet).

	To detect the stimulated Hawking process, we sent shallow water waves
toward the effective white hole horizon, which sits on the lee side of the
obstacle. We conducted a series of experiments, with q = 0.045 ${\rm m}^2$/s and h =
0.194 m, and examined 9 different ingoing frequencies between 0.02 and 0.67
Hz, with corresponding still water wavelengths between 69 and 2.1 meters. This
surface was imaged at 20 frames per second, for about 200 s. In all cases we
analysed a period of time which was an exact multiple of the period of the
ingoing wave, allowing us to carry out sharp temporal frequency filtering of
the signals (i.e., eliminating spectral leakage).
\bea
\xi=\int_0 {dx\over v(x)}
\eea

The analysis of the surface wave data was facilitated by introducing the
convective derivative operator $\partial_t + v(x) \partial_x$. We redefine the spatial
coordinate using, where $x$ is the distance downstream from the right hand edge
of the flat portion of the obstacle. The $\xi$ coordinate has dimensions of time,
and its associated wave number $\kappa$ has units of Hz. The convective derivative
becomes $\partial_t + \partial_\xi$, or, in Fourier transform space,  $f +
\kappa$. This is the term that
enters the conserved norm.  From equations (35), (36) 
 and (87) of reference [9] we find that the conserved norm has the form 
\bea
\int{{\vert A(f,\kappa)\vert^2\over f+\kappa} d\kappa}
\eea
where $A(f,\kappa)$ is the $t,~\xi$ Fourier transform of the vertical
displacement of the wave. In using this coordinate system the outgoing waves
have an almost uniform wavelength even over the obstacle slope.

\section{Results}

We will illustrate the pair-wave creation process by presenting the results
for $f_{in} = 0.185$ Hz. In this case, we analyzed images from exactly 18 cycles,
measuring the free surface along approximately 2 m of the flow including the
obstacle. After converting to $\xi$-coordinates (6), we calculated the
two-dimensional Fourier transformation as displayed in Fig.\ 4a. Note that the
amplitudes of the Fourier transform at frequencies above and below 0.185 Hz
are very small, indicating that the noise level is small

As expected, there are three peaks, one corresponding to the ingoing shallow
water wavelength around $\xi = 0$, and the other two corresponding to converted
deep water waves peaked near $\kappa^+_{out} = 9.7$ Hz and $\kappa^-_{out} =
-10.5$ Hz. The former
is a positive norm and the latter a negative norm outgoing wave, see equation
(7). 

\begin{figure}
\begin{center}
\includegraphics[width=80mm]{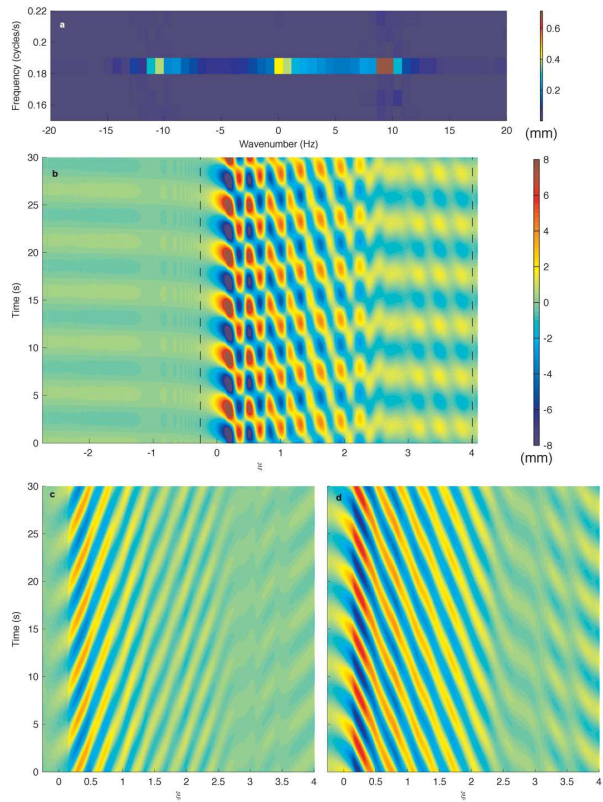}
\caption{{\bf  | Pair-wave creation.} Demonstration of pair-wave conversion
of an ingoing frequency of 0.185 Hz: {\bf a}, Fourier transform of unfiltered wave
characteristic. {\bf b}, Filtered wave characteristic containing only the ingoing
frequency band. {\bf c} and {\bf d}, wave characteristics for filtered negative and
positve norm modes. (The colours represent the amplitudes of the waves, see
colour bars.)
}
\end{center}
\end{figure}

In Fig.\ 4b we plot the wave characteristics (amplitude as function of t and
$\xi$)
filtered to give only the temporal 0.185 Hz band. Figures 4c and 4d are the
characteristic plots where we further filter to include only $\kappa < -1$ Hz
and $\kappa >
1$ Hz respectively. These are the negative and positive norm outgoing
components without the central peak of the ingoing wave (because of their very
long wavelengths and the rapid change in wavelength as they ascend the slope,
the incoming waves have a very broad Fourier transform). Recall that since we
are only interested in counter-propagating waves, we defined positive phase
and group speeds as pointing to the left. As expected from the dispersion
relationship, see Fig.\ (1), the negative norm waves have negative phase
velocity, while the positive norm waves have positive phase velocity. The
complex structure in the characteristics of Fig.\ 4b arises because of the
interference between the three components, the original ingoing wave, and the
positive and negative norm outgoing waves. In Fig.\ 4b, we see that the ingoing
wave is blocked around $\xi= 0$, with only a small component penetrating into the
region over the top of the obstacle $\xi < 0$.

Our key results are presented in Fig.\ 5. Figure 5a shows the amplitude of the
spatial Fourier transform at three selected ingoing frequencies. As the
frequency increases, the ratio of the negative norm peak to positive norm peak
decreases. Furthermore, the location of the positive norm peak moves slightly
toward zero as the frequency increases, while the negative norm peak moves
away from zero. This is to be expected from the location of the allowed
spatial wavenumber from the dispersion plot, see Fig.\ 1. The red curve in
Fig.\ 5a
 shows the Fourier transform in the adjacent temporal frequency bands for
the sample case of 0.185 Hz. This is a representation of the noise, and is a
factor of at least 10 lower than the signal in the 0.185 frequency band

\begin{figure}
\begin{center}
\subfigure{
\includegraphics[width=80mm]{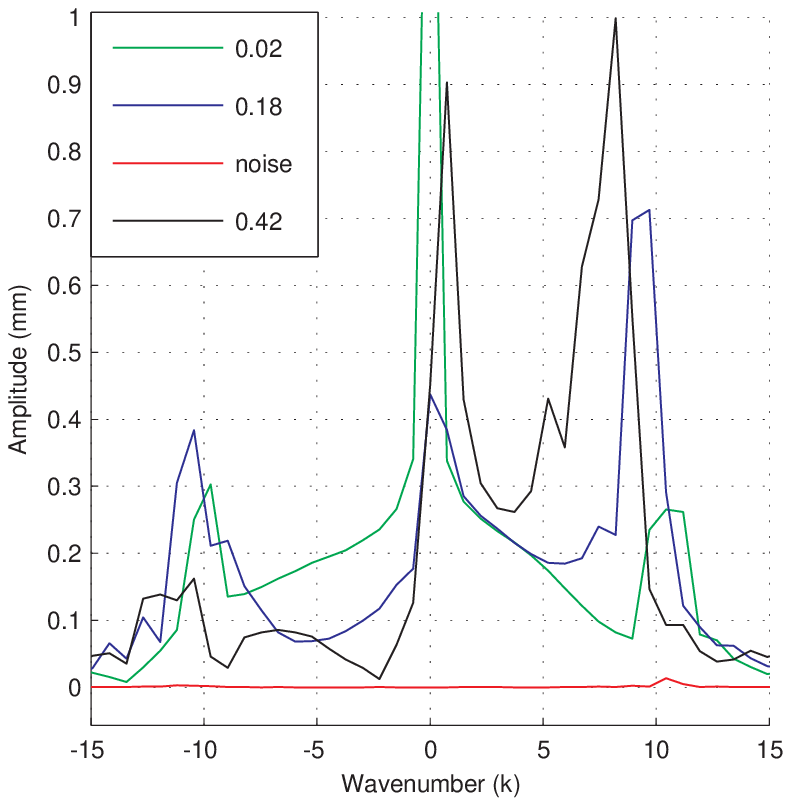}
}
\subfigure{
\includegraphics[width=80mm]{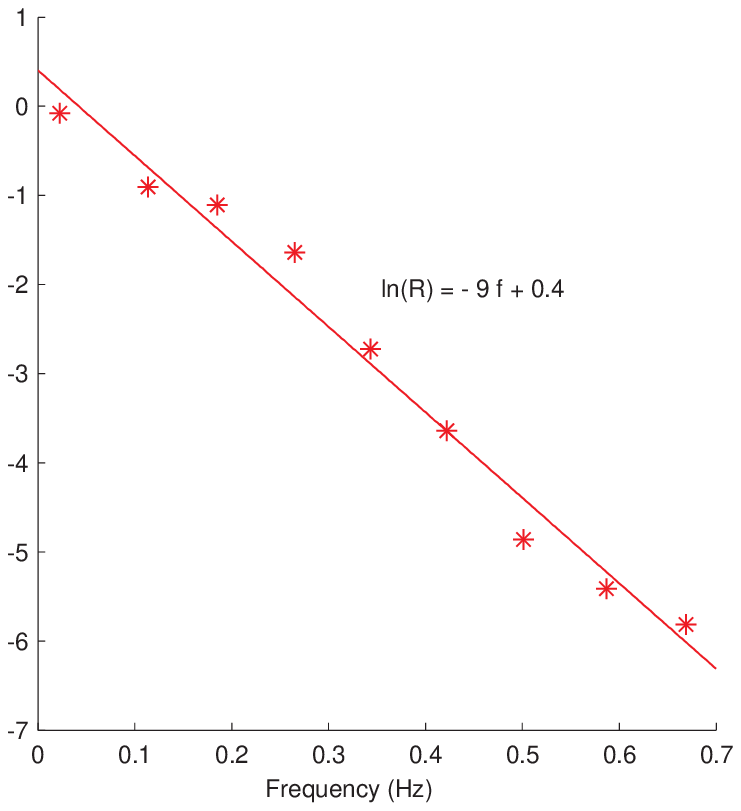}}

\caption{{\bf | Amplitudes and thermal spectrum. a}, Absolute value of
three different ingoing frequency bands, and typical noise level (red line).
{\bf b}, Natural log ratio R of negative and positive norm components
(Eqn.(1)) in between 0.02 and
0.67 Hz (red stars), and linear least-squares fit (red line).
}
\end{center}
\end{figure}

To test whether or not the negative norm wave creation was due to
non-linearities we repeated the runs at all frequencies with 50\% larger
amplitudes. The converted wave amplitudes did, in fact, scale linearly.  

The crucial question is: Does the ratio of the negative to positive norm
outgoing waves scale as predicted by the thermal hypothesis of equation (1)?
This is shown to be the case in Fig.\ 5b, where the norm ratios are plotted as
a function of ingoing frequency. To calculate the norm of the outgoing waves
we integrate $\int \vert A(f,\kappa)\vert^2/(f+\kappa) d\kappa$ over 
the peaks to evaluate $\vert \alpha_f\vert^2$ or $\vert \beta_f\vert^2$
. In Fig.\ 5b the points represent the natural  log of the ratios of these areas for
each of the input frequencies we tested. The thermal hypothesis is strongly
supported, with linear regression giving an inverse slope of 0.11 Hz and an
offset close to zero. The slope corresponds to a temperature of $T = 5 \cdot
10^{-12}$ K.

We see from Fig.\ 4c,d that the region of ``wave blocking" where the ingoing
wave is converted to a pair of outgoing waves, is not a phase velocity horizon
(where the phase velocity in the laboratory frame goes to zero). This is true
even for the very lowest frequencies.  The usual derivation of the temperature
from the surface gravity relies on this conversion occurring at a phase
velocity horizon. This makes the calculation of the surface gravity, and thus
the predicted temperature uncertain. In our case estimates of the surface
gravity give a predicted temperature of the same order as the measured
temperature. What is important is that the conversion process does exhibit the
thermal form predicted for the Hawking process. 

\section{Summary}

We have conducted a series of experiments to verify the stimulated Hawking
process at a white hole horizon in a fluid analogue gravity system. These
experiments demonstrate that the pair-wave creation is described by a
Boltzmann-distribution, indicating that the thermal emission process is a
generic phenomenon. It survives fluid-dynamical properties, such as turbulence
and viscosity that, while present in our system, are not included when
deriving the analogy. The ratio is thermal despite the different
dispersion relation from that used by Hawking in his black hole derivation.
This increases our trust in the ultraviolet independence of the effect, and
our belief that the effect depends only on the low frequency, long wavelength
aspects of the physics. When the thermal emission was originally discovered by
Hawking, it was believed to be a feature peculiar to black holes. Our
experiments, and prior numerical work [6,12], demonstrate that this phenomenon
seems to be ubiquitous, and not something that relies on quantum gravity or
Planck-scale physics. 

While our experiments measure only the stimulated emission from this white
hole analogue, it has been known since Einstein's work [23] that there is a
very close relation between spontaneous and stimulated emission from a quantum
system. Furthermore the time reversal invariance of the theory leads to the
equivalence of black and white hole horizons. It would still be exciting to
measure the spontaneous emission from a black hole. While finding small black
holes to test the prediction directly is beyond experimental reach, such
measurements might be achievable in other analogue models, like Bose Einstein
condensates, or optical fibre systems [24-28].

\acknowledgments{
We thank Mauricio Richartz for his help during the initial stages of this
project. WGU and GAL thank the Natural Sciences and Engineering Research
Council for grants which supported this work. WGU thanks the Canadian
Institute for Advanced Research, and GAL thanks the Canada Research Chairs
program, for their support of this research. SW was supported by a Marie Curie
Fellowship EMERGENT- 2007-SW. We thank Piyush Jain and Ralf Schützhold for
comments on the final manuscript and Trevor Guerard for help with fig 2. We thank the Department of Civil Engineering
for the use of the flume and experimental space, which displaced undergraduate
teaching. Their willingness to make do with other equipment and space made our
experiments possible. 
}

\end{document}